# SURPRISING RESULTS FROM COSMIC RAYS

G. Wilk[†] and Z. Włodarczyk[‡]

[†]*Soltan Institute for Nuclear Studies, Hoża 69, PL-00-681 Warsaw, Poland*
[‡]*Institute of Physics, Pedagogical University, Leśna 16, PL-25-509 Kielce, Poland*

# SURPISING RESULTS FROM COSMIC RAYS [1]

G. Wilk[1],[†] and Z. Wlodarczyk[2],[‡]

[1]*Soltan Institute for Nuclear Studies, Warsaw, Poland*
[2]*Institute for Physics, Pedagogical University, Kielce, Poland*

A number of seemingly *exotic* phenomena seen in the highest cosmic-ray experiments are briefly discussed. We argue that they indicate existence of non-statistical fluctuations and strong correlations in the fragmentation region of multiparticle production processes unaccessible to the present accelerators.

**1. Introduction.** Physics has its roots and final justification in experiments. High energy collision experiments, which we are interested in here, originated with discovery (in 1912) of cosmic rays, CR. For more than half of century CR were the only source of high energy collisions and only relatively recently accelerators reached comparable energies of the order of TeV in cms (Fermilab). It is possible, however, that the L(arge) H(adronic) C(ollider) planned at CERN will be the last big accelerator constructed (the construction of even bigger machine, SSC in USA, which was planned to reach energy 40 TeV in cms, has been already stopped). In that case CR would again remain the sole source of our experimental information concerning interactions of *extremaly high energy* [2] It is therefore interesting what one can expect from CR and this was the main reason of our presentation at this otherwise highly theoretical meeting. The other reason is that CR turn out increasingly important also at present accelerator energies because they provide information complementary to that obtained from accelerators [3].

The history of CR is full of apparent *anomalies*, i.e., results which where extremaly surprising (i.e., with no immediate explanations) at the time of their registration. Some turned out to be experimetal artifacts but those which survived additional tests were all confirmed in accelerator experiments at later time. For example, *positron*, $e^+e^-$ pair production and *electromagnetic cascades*, *muon*, $\tau$ and $\pi$ mesons were all first observed in CR before 1950! This list continues with strange particles and hyperfragments and ends with charm observed already in 1971 by Japanese group and rejected as fake event at that time. One can add also the growth with energy $s$ of such characteristics of collisions as total inelastic cross section $\sigma_{inel}^{tot}$, mean multiplicity $\langle n(s) \rangle$ and mean transverse momenta of produced secondaries $\langle p_T \rangle$. The weak point of CR is the procedure leading from the raw data to final conclusions which is usually extremaly involved and tedious [1]. However, if CR are really going to be our sole future source of information on high energy production processes, we should develop methods to check and interpret their results properly without further help of accelerator data rather than dismiss them lightly whenever we do not understand them

---

[2]It is worth to realize that even CR have their natural limit due to the microwave background radiation filling the Universe which practically destrois everything with energy greater than $\sim 10^{20}$ eV ($\sim 1000$ TeV in cms) on distance up to 100 Mpc.

[3]Accelerators measure a big (and growing with energy) number of relatively low energetic secondaries produced mostly in the central region of collision whereas CR measure the most energetic fragments of the projectiles, i.e., they are sensitive to the fragmentation region of high energy collisions including diffraction dissociation processes.

at a given moment.

## 2. Anomalous CR results today.

There is quite a number of CR events observed at present and regarded as anomalous (or exotic, i.e., without established and sound explanation at present). Referring to [2] for more detailed description (and definitions) we shall just list them here: (*i*) *Centauros* - events in which one observes $\sim 70-80$ charged hadrons without any $\pi^0$'s; (*ii*) *Chirons* - events in which one registers families with extraordinary large lateral spread; (*iii*) *Colimated shower clusters* - i.e., families with a very colimated center, spreading only a few milimiters in diameter in the X-ray film where they are registered; (*iv*) *Penetrating cascades* (more than electromagnetic ones) and at the same time events with mean free path of the shower-initiating particle being only $\frac{1}{2} - \frac{1}{3}$ of the geometrical mean free path of the nucleon (which is the most abundant component of the original CR); (*v*) *Coplanarity of multi-halos* - phenomenon of the alingment of structural objects seen in the families in the plane of target diagram; (*vi*) *Long-flying component* - phenomenon of the aparent *non-exponential* absorption of hadrons at large depths of the emulsion chambers. Instead of describing them one by one, for which we have no space here, we shall try to summarize what they mean and than to provide our possible explanations of at least some of them.

We concentrate on events called *families* (forming centers of the so called Extensive Air Showers (EAS)) observed in the emulsion chambers (EC) exposed on mountain altitudes of Chacaltaya(5220m) and Pamir(4370m) [3] and Kanbala(5500m) and Fuji(3770m) [4] (estimated energy involved is of the order of the present CERN SPS and Fermilab Tevatron colliders and above). What was said above can be, for our further purpose, summarized in the following way. Family results from some primary interaction by impinging projectile the type of which and energy is essentially unknown (one knows only initial composition and spectrum [1,3,4]). This interaction usually occurs at some height $H$ above the EC which can be estimated either by triangulation from the tracts registered in EC or from the relation: $E \cdot R = p_T \cdot H$, where $E$ denotes measured energy of cascade, $R$ its (measured) distance from the centre of the family and $p_T$ (assumed) mean transverse momentum of secondaries. After first interaction it develops due to the next-generation interactions[4] and finally EC registers a number of tracts which can be divided into electromagnetic (gamma component or gamma families) and hadronic (hadronic component or hadronic families) in origin and have distinctive behaviour. Usually they are grouped in kind of bundles (showers) entering points of which form some distinct patter of spots of different diameters and optical densities (darkness) in the X-ray films separating emultion plates. What and how is really measured depends on the construction details of the EC used [3,4]. The picture which emerges and which visualises and summarizes somehow what was listed above is that sometimes we register unusual (nonstatistical) fluctuations in gamma/hadron ratios observed (*Centauro, Mini-Centauro* and *Anti-Centauro* events), sometimes observed spots are very much concentrated (signaling narrow families and very strong energy concentration) and sometimes their lateral spread is very large (all these above of what one could expect as being statistical fluctuation if the cascading proces of formation in the atmosphere of a given family). On the other hand one observes also that the structure of families seen by EC is not uniform, it displays sometimes azimuthal asymmetry and some alingnment of registered showers (coplanarity). One observes also such phenomenon that shower entering EC starts its interaction there very soon (much faster than expected from the known hadronic cross section for any interaction), i.e., its mean free path is strangely small (indicating naively some "large" object flying by) whereas at the same time this shower continues to interact very deeply in the EC (i.e., it penetrates very deep as if only weakly interacting on the way).

---

[4]This is described by the classical diffusion like equation where elementary interactions enter through their respective cross sections and inelasticity coeficients, cf. for example [5].

**3. What does it mean?**   There is alway an immediate question: does all this signal *new physics* or can it be expained in more orthodox way? We are iclined to argue that it this the later which should be pursued[5]. The way of reasoning is following: because the family events are sensitive to the fragmentation/diffractive dissociation (DD) region of the phase space for elementary collisions (cf. Fig. 1 [6]), which is essentially not explored for this energy range, all extrapolations of used phenomenological parametrizations can be misleading to an unknown extent. It was already shown in a number of papers that inclusion of diffractive component in terms of the diffractive production of very high masses can indeed explain to some extend both *Centauros* and some alingment problems as well. Such obvious candidates as QCD jets and possible quark compositeness failed to describe data alone [7].

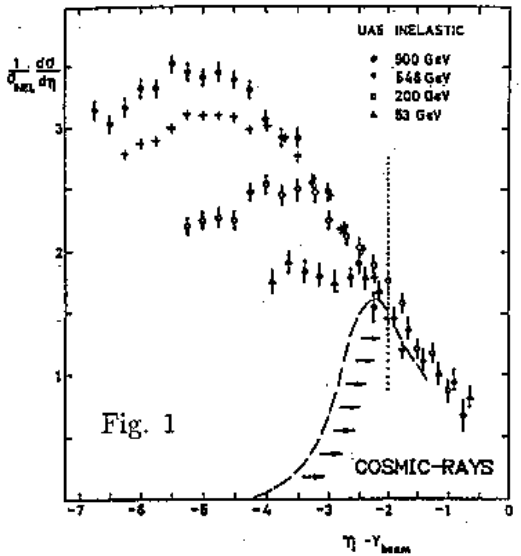

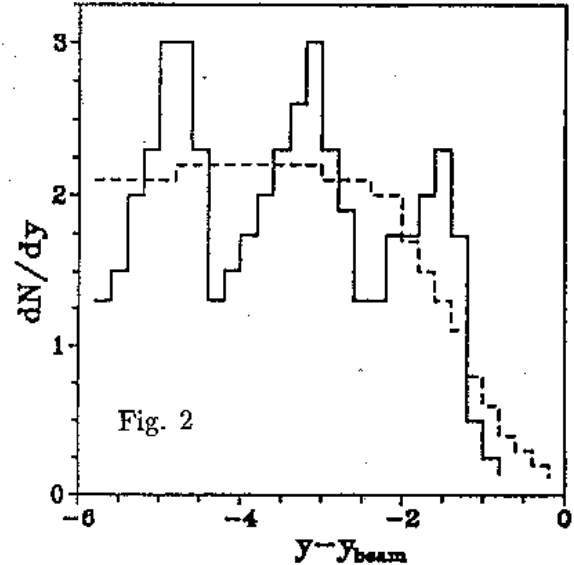

Here we would like to bring ones attention to the fact that there is still one important feature of high energy production processes which is observed and widely studied at accelerator energies (albeit in central region only) but so far has not yet been explored in the analysis of the CR, in particular to data from families we are interested in. This feature is the presence of fluctuations and correlations in the multiparticle production processes visualising themselves as bunching of particles in the phase space called intermittency and correlations between identical particles known as Bose-Einstein correlations (BE) [8].

---

[5]We have already some experience in that field showing that apparently very strong signal of intermittency in elementary collisions deduced from CR comes in reality from the (highly stochastic but natural) development of families itself and that long-flying component mentioned before is just reflection of the possible fluctuations in the cross-sections observed already at accelerators [6].

[6]Here pseudorapidity distributions of charged particles measured in UA5 experiment at different energies at CERN SPS accelerator are shown together with the particles selected in gamma families with total energy $\sum E_\gamma \geq 100$ TeV.

Indeed, introducing even very simple and obviously preliminary bunching of particles produced in elementary collisions in the phase space changes already substantially final results. As first example let us mention that local fluctuations in rapidity density (such as presented in Fig. 2 by full line, i.e., not changing the mean rapidity distribution presented by dashed line), which is a classical example of intermittency, changes the (so called rejuvenated [9]) mean multiplicity of gamma families $\langle n' \rangle$ by more than 10% [7]. If we perform such bunching in all 3 dimensions by arranging identical particles (like-charged $\pi$'s) in such a way as to minimize their mutual momenta $|\vec{p}_i - \vec{p}_j|$ in each cluster it turns out that we can describe azimuthal asymmetry of a family[8] [9] and (rejuvenated) mean multiplicity mentioned above at the same time. This is best seen in Fig. 3 [9].

In the same way we can explain phenomenon that in some hadronic families there occurs an unexpectedly large fraction of energetic hadrons, $\langle N \rangle = 1.35 \pm 0.08$, carrying fraction

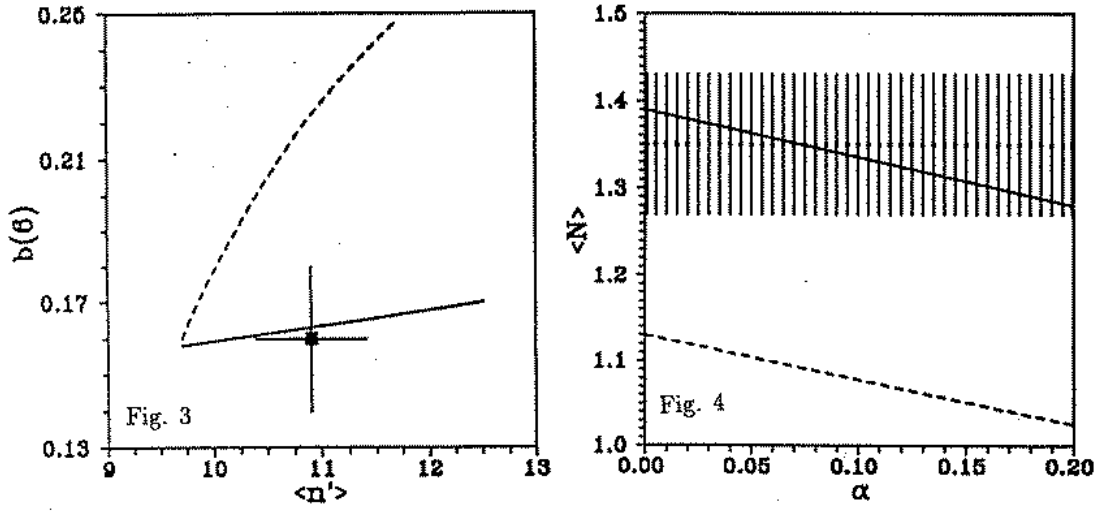

$\langle Q \rangle = 0.51 \pm 0.03$ of the energy of the whole family each. It is sensitive to the violation of the Feynman scaling, however, so far models without bunching were not able to describe it, cf. Figs. 4 and 5 [10].

---

[7] Increasing it from $\sim 10$ as given by so called scaling violation models to $\sim 11$ which allows for smaller scaling violation apparently seen in CR data. The so called Feynman scaling notion means independence of the scaled momenta of produced particles on the collision energy and its presence or absence is a subject of debate at present.

[8] Defined as: $b = \sum E_i Y_i^2 / \sum E_i X_i^2$ with $E_i$ being energy of $i^{th}$ particle and $X$ axis is chosen such as to minimize $\sum E_i Y_i^2$.

[9] Here the azimuthal-asymmetry coefficient $b(6)$ for the six highest energy particles is presented as a function of $\langle n' \rangle$ for rejuvenated families with relative threshold $f'_{min} = 0.04$ and the visible energy registered exceeds 100 TeV; the experimental point [9] can be reached only if correlations (in the form descibed above) are included (solid line), uncorrelated particle production (dashed line) is clearly excluded.

[10] Here $\langle N \rangle$, Fig. 4, and $\langle Q \rangle$, Fig. 5, are plotted as a function of scaling violation degree $\alpha$ (for inclusive

As the last example we shall mention here the anomalous transition behaviour of showers of hadronic origin observed at Chacaltaya EC [3]. As we have already mentioned at the same time we observe small mean free path and strongly penetrating showers. It is seen as a number of sub-showers which maxima can be observed in EC. The immediate explanation is a cluster of particles with very small transverse momenta $p_T$ (of the order of a few tens of MeV/c only!) propagating through the EC, cf. Fig. 6 [11]. This is perhaps most spectacular example showing the necessity of accounting for correlation phenomena also in CR events. But it opens another box of unknowns as there is an obvious question how to produce such big number of strongly correlated particles. The immediate idea is to look more closely for fragmentation/DD properties of multiparticle production processes. It means, however, entering uncharted teritory as essentially nothing is known about BE correlations in the fragmentation or DD regions of production processes except of the undirect expectations that there should be no such effect at all. The reason is that on the basis of the quantum statistical (QS) approach to multiparticle production processes it was argued that the particles in the fragmentation region should be produced in the *coherent* fashion for which BE correlation are zero [10]. On the other hand, if particles are produced through decays of the colour field strings [11] one can argue that in the fragmentation region one should observe at least correlations similar to those seen in the $e^+e^-$ annihilation processes in which single string configuration is dominant. Because BE correlation do exist there in their full strength spectra the mean value of Feynman variable $x$ which in CR is taken as a fraction of energy in respect to the total measured energy, this is what rejuvenation mentioned before means, decreases as $\langle x \rangle \sim s^{-\alpha}$). Again uncorrelated particle models (dashed line) are completely ruled out whereas those with bunching (modeled by assuming that particles are produced in pairs with close values of their momenta) allow for some degree of scaling violation.

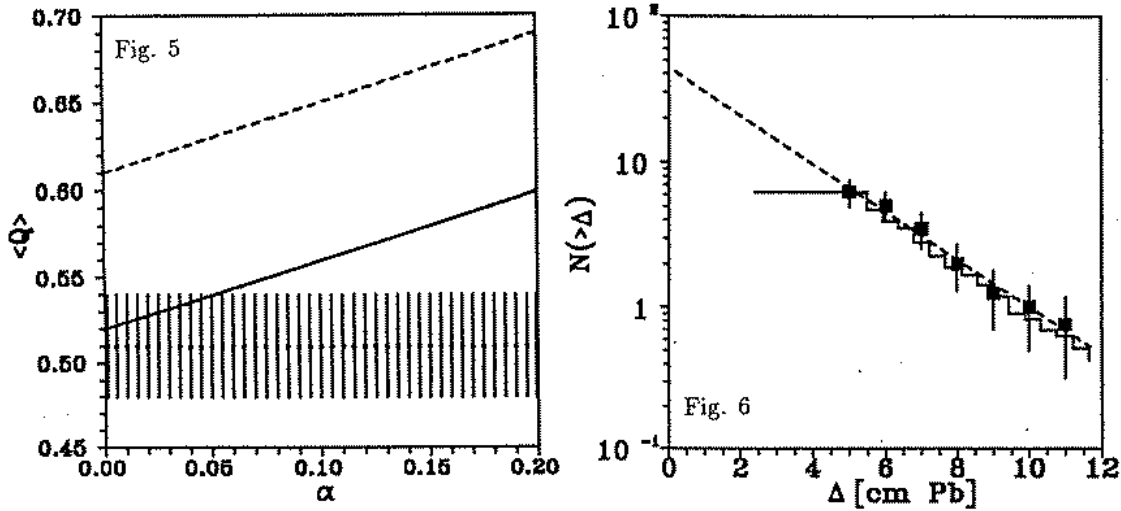

[11] Here we plot data points for distribution of distances $\Delta = t_{i+1} - t_i$ between the sub-shower maxima positions $t_i$ as measured for 4 events from Chacaltaya EC [3] ($\langle \Delta \rangle = 7.58$ cm Pb and the observed number of sub-showers $\langle N \rangle = 7.52$). Solid histogram corresponds to our results assuming propagation of bunch of 45 pions (which number was deduced from the extrapolation of data to $\Delta = 0$; mean free path $\lambda = 250\,\mathrm{g/cm^2} = 22$ cm Pb; showers initiated at distances smaller than $0.24\,\lambda$ are not resolved and those separated by more than $0.7\,\lambda$ are excluded from analysis).

[12] (what is still unsolved surprise of strong interactions) one is than, indeed, expecting at least some BE correlations operating also in the case of CR events under considerations [12].

**4. Conclusions.** We presented here a survey of apparently surprising results from CR and demostrated on a couple of examples that most probably they do not need any new physics but only a more clever use of dynamical ideas already known from accelerator experiments. On the other hand it seems even more clear that one can also learn from the CR experiments on some yet unknown features of the fragmentation/DD processes in multiparticle production which are not directly accessible for accelerators. Results shown here must be regarded as preliminary only and showing a possible subject for further investigations [2].

---

[12]It should be remembered, however, that BE correlations in $e^+e^-$ annihilations were measured so far only in the central region of string(s); therefore our statement must be still regarded as a hypothesis only.